\documentclass[pre,onecolumn,amsmath,draft,superscriptaddress,floatfix]{revtex4}

\usepackage[final]{graphicx}

\begin{document}

\title{Generalized entropy arising from a distribution of  $q$-indices}

\author{G. A. Tsekouras}
\email{gtsek@chem.demokritos.gr}
\affiliation{Institute of Physical Chemistry, National Center for Scientific Research
``Demokritos'', 15310 Athens, Greece}
\affiliation{Department of Physics, University of Athens, 10679 Athens, Greece}
\author{Constantino Tsallis}
\email{tsallis@cbpf.br, tsallis@santafe.edu}
\affiliation{Santa Fe Institute, 1399 Hyde Park Road,
Santa Fe, New Mexico 87501,  USA}
\affiliation{Centro Brasileiro de Pesquisas F\'isicas, Rua Xavier Sigaud 150,
22290-180 Rio de Janeiro-RJ, Brazil}

\begin{abstract}

It is by now well known  that the Boltzmann-Gibbs (BG) entropy $S_{BG}=-k\sum_{i=1}^W p_i \ln p_i$ can be usefully generalized into the entropy  $S_q=k\,(1-\sum_{i=1}^Wp_i^{\,q})\,/\,(q-1)$ ($q\in \mathcal{R}; \;S_1=S_{BG}$). Microscopic dynamics determines, given classes of initial conditions, the occupation of the accessible phase space (or of a symmetry-determined nonzero-measure part of it), which in turn appears to determine the entropic form to be used. This occupation might be a uniform one (the usual {\it equal probability hypothesis} of BG statistical mechanics), which corresponds to $q=1$; it might be a free-scale occupancy, which appears to correspond to $q \ne 1$. Since occupancies of phase space more complex than these are surely possible in both natural and artificial systems, the task of further generalizing the entropy appears as a desirable one, and has in fact been already undertaken in the literature. To illustrate the approach,  we introduce here a quite general entropy based on a distribution of $q$-indices thus generalizing $S_q$. We establish some general mathematical properties for the new entropic functional and explore some examples.
We also exhibit a procedure for finding, given any entropic functional, the $q$-indices distribution that produces it. Finally, on the road to establishing a quite general statistical mechanics, we briefly address possible generalized constraints under which the present entropy could be extremized, in order to produce canonical-ensemble-like stationary-state distributions for Hamiltonian systems.  
\end{abstract}

\maketitle

\section{Introduction}

Since Gibbs' pioneering words \cite{Gibbs}, it has become well established that standard, Boltzmann-Gibbs (BG) statistical mechanics and the associated thermodynamics are valid when certain conditions are satisfied. The typical situation occurs for microscopic dynamics exhibiting {\it strong chaos} (i.e., positive largest Liapunov exponent) and, consistently, exponential mixing, ergodicity, short relaxation times, Euclidean-like occupation of phase space and usual thermodynamic extensivity. This is the scenario which typically occurs for short-range-interacting many-body Hamiltonian systems, even if,  rigorously speaking, the strict necessary and sufficient conditions remain to be proved.  On the other hand, a vast class of natural and artificial systems exists for which the largest Liapunov exponent vanishes, situation which is referred to as {\it weak chaos}. Weak chaos is typically associated with power-law, instead of exponential, sensitivity to the initial conditions and relaxations, (multi)fractal occupation of phase space and thermodynamic nonextensivity. 

During the last 15 years a great deal of attention has been focused on systems where the usual BG statistical mechanical concepts prove inadequate. Phenomena which exhibit such anomalous behavior occur in systems involving long-range interactions (e.g., gravitation), spin-glasses, manganites, long-range micro- or mesoscopic memory, turbulence in nonneutral plasma and classical fluids, L\'evy anomalous diffusion, granular systems, phonon-electron anomalous thermalization in ion-bombarded solids, solar neutrinos, cosmic rays, galactic peculiar velocities, cosmological systems, high-energy collisions of particles, black holes, quantum entanglement, lattice Lotka-Volterra growth model, {\it Hydra viridissima}, financial indices and others (see \cite{GellmannTsallis04,SwinneyTsallis04,HerrmannBarbosaCurado04} for recent reviews). 

In 1988, a possible generalization of BG statistical mechanics was proposed \cite{Tsallis88} on the basis of the following entropy
\begin{equation}
S_q = k \frac{1-\sum_{i=1}^W p_i^q}{q-1}  \;\;\;\;(q \in \mathcal{R}; \; S_1 \equiv S_{BG}=-k \sum_{i=1}^W p_i \ln p_i) \, , 
\end{equation}
where $k$ is a positive constant, usually Boltzmann constant, but from now on taken to be unity for simplicity. This generalization of BG statistical mechanics is usually referred to as {\it nonextensive statistical mechanics}. Its connection with thermodynamics was established in \cite{CuradoTsallis91} and later redefined in \cite{TsallisMendesPlastino98}. Its denomination {\it nonextensive} comes from the following property: if we have two probabilistically {\it independent} systems $A$ and $B$, i.e., $p_{ij}(A+B)=p_i(A)p_j(B)$, we straightforwardly verify that
 \begin{equation}
S_q(A+B) = S_q(A)+S_q(B)+(1-q)S_q(A)S_q(B) \;. 
\end{equation}
Consequently, $q=1$ (the BG case) corresponds to extensivity, whereas $q<1$ ($q>1$) corresponds to superextensivity (subextensivity), where the nonnegativity of $S_q$ has been taken into account. We should stress at this point that the denomination ``nonextensive" can be misleading. Indeed, when special correlations are addressed (instead of independent probabilities), it has been shown \cite{TsallisSato} that $S_q$ is {\it extensive} for the appropriate value of $q$. When no scale-invariant (or similar) correlations are present, the appropriate value of $q$ equals unity.

Within nonextensive statistical mechanics, many of the above cited anomalous systems \cite{GellmannTsallis04,SwinneyTsallis04,HerrmannBarbosaCurado04} have found a frame of interpretation. The success of such type of approach has led to even more general formalisms. Such is the case of Beck-Cohen superstatistics \cite{BeckCohen03,Beck04,TsallisSouza03,SouzaTsallis03}, based on fluctuations of parameters such as the temperature \cite{WilkWlodarczyk00,Beck01}.  Other examples, such as the early proposals by Curado\cite{Curado} and by Anteneodo and Plastino \cite{AnteneodoPlastino}, as well as, more recently,
Kaniadakis et al formalism based in the $\kappa$-entropy \cite{Kaniadakis}, do exist \cite{Naudts,SouzaTsallis04} that have started to exhibit their usefulness for applications. 

In this spirit we propose here a self consistent and hopefully applicable generalization of standard non-extensive statistical mechanics. We will base our effort on the idea of using not just one $q$ entropic index but a whole distribution of them. The starting point, successfully applied in re-association of folded proteins \cite{TsallisBemskiMendes99}, cosmic rays \cite{TsallisAnjosBorges03} and economics \cite{Borges04}, is an ordinary differential equation whose solution exhibits a crossover from a $q$-statistics to a $q^\prime$-statistics. This solution might be thought as a generalized ``exponential", whose inverse function is therefore a generalized ``logarithm". Two natural manners appear then for extending this type of crossover to a whole spectrum of $q$'s, the solution of the full-spectrum ordinary differential equation being interpreted as a quite general exponential form, whose inverse is therefore a quite general logarithm. The first manner consists in using the generalized logarithm for defining a generalized entropy which contains $S_q$ as a particular case; from there, as usual, we obtain the stationary-state distribution by optimizing the generalized entropy under appropriate constraints. This distribution does {\it not} necessarily have the generalized exponential form, a nice property which however does occur for nonextensive statistical mechanics. The second manner is to look for a generalized entropy which, when optimized under appropriate constraints, precisely provides the given generalized exponential. Such entropy will in general {\it not} be based on the corresponding generalized logarithm, a nice property which, as already said, does occur for nonextensive statistical mechanics. The first of these two manners is, although far from trivial, mathematically simpler. This is the one to which the present paper is dedicated. The second manner remains as a task to be undertaken in a later occasion.  

In Section II we introduce our formalism by generalizing the logarithmic and exponential functions, as well as the corresponding entropic functional. In Section III we study some of their basic properties. In Section IV we illustrate the theory with some nontrivial examples, namely a Gaussian distribution of q-exponents and a sum of two delta functions. In Section V we introduce and work out the idea of the inverse transformation which enables to find, for a
given entropic functional, the distribution of exponents that produces it; an illustration is provided as well. 
In Section VI we address the problem of the constraints that can be used in order to obtain a canonical-ensemble-like stationary distribution. In Section VII we investigate the relation between the present formalism and Beck-Cohen superstatistics.  Finally, in Section VIII, we recapitulate our results and discuss some future perspectives.

\section{Generalizing the nonextensive entropy functional}

As well known, BG statistical mechanics is based on the entropy functional
\begin{equation}
S_{BG}=\sum_{i=1}^W p_i \ln \frac{1}{p_i}
\end{equation}
This is a convenient manner of writing $S_{BG}$. Indeed, at equiprobability we have $p_i=1/W$, hence the functional takes the value $k \ln W$ since the number of (equal) terms precisely cancels  $p_i$ in front of the logarithm.
The $\ln x$ function is the inverse of the $\exp x$ function, which is the solution of the following differential equation \cite{GellmannTsallis04,SwinneyTsallis04}:
\begin{subequations}
\begin{equation}
\frac{dy}{dx}=y
\end{equation}
\begin{equation}
y(0)=1
\end{equation}
\end{subequations}

In the frame of nonextensive statistical mechanics Eqs. (3) and (4a) are generalized as
follows (Eq. (4b) is maintained as it stands):
\begin{equation}
S_q=\sum_{i=1}^W p_i \ln_q \frac{1}{p_i}
\end{equation}
and
\begin{subequations}
\begin{equation}
\frac{dy}{dx}=y^q
\end{equation}
\end{subequations}

The $\ln_q x\equiv [x^{1-q}-1]/(1-q)$ function is the inverse of the $\exp_q x \equiv [1+(1-q)x]^{1/(1-q)}$ function, which is the solution of Eq. (6a). 
For $q=1$ we have $\ln_1x=\ln x$ and $\exp_1 x = \exp x$, consequently we fall back to the BG case. For $q<1$, $exp_q \,x$ is taken to be zero if $1+(1-q)x \le 0$. 

Following the above rational used in generalizing BG statistics to nonextensive
statistics we can go further and examine the effect of the contribution of a
whole distribution of $q$ exponents. Within the present generalization the entropy is assumed to be of the following form:
\begin{equation}
S_{\{f\}}=\sum_{i=1}^W p_i \ln_{\{f\}} \frac{1}{p_i} \equiv           \sum_{i=1}^W s_{\{f\}}       \;,
\end{equation}
where the $\ln_{\{f\}}x$ function is the inverse of the $\exp_{\{f\}}x$ function, defined as the solution of
\begin{subequations}
\begin{equation}
\frac{dy}{dx}=\int_{-\infty}^{+\infty}f(\tau)\,y^\tau d\tau
\end{equation}
\begin{equation}
y(0)=1
\end{equation}
\end{subequations}
with
\begin{equation}
\int_{-\infty}^{+\infty} f(\tau) \,d\tau=1\;\;\; (f(\tau)\geq 0, \, \forall \tau \in \mathcal{R}).
\end{equation}
We will call the nonnegative, normalized distribution $f$ the {\it q-spectral function (QSF)} since it
represents the spectrum of $q$ entropic indices that contribute to the entropic functional.
In this generalized framework, for $f(\tau)=\delta(\tau-1)$ we recover the BG
entropy $S_{BG}$ and for $f(\tau)=\delta(\tau-q)$ we recover the nonextensive entropy $S_q$, where $\delta$ denotes Dirac's distribution. In fact, it is not necessary to impose normalization onto $f(\tau)$; it is enough to only ask for the integral $\int_{-\infty}^{+\infty} f(\tau) \,d\tau$ to be {\it finite}. If the function $f(\tau)$ is not normalized, it introduces a very simple modification, namely it multiplies the associated generalized logarithm by a constant \cite{remark}. It is therefore just for simplicity that we hereafter demand $f(\tau)$ to be normalized. 

We will now try to establish the form of $\ln_{\{f\}} x$ . By definition of the $\exp_{\{f\}}x$ function, it is
\begin{equation}
\frac{d\left [ \exp_{\{f\}}x \right ]}{dx}=\int_{-\infty}^{+\infty} f(\tau) \left[ \exp_{\{f\}}x \right ] ^\tau d\tau \;,
\end{equation}
hence, by setting $x=\ln_{\{f\}}y$ ,
\begin{equation}
\frac{dy}{d\left [ \ln_{\{f\}}y \right ]}=\int_{-\infty}^{+\infty} f(\tau) \,y^\tau d\tau \;, 
\end{equation}
hence
\begin{equation}\label{eqln}
\ln_{\{f\}}x=\int_1^x \left \{ \int_{-\infty}^{+\infty} f(\tau) \,u^\tau d\tau
\right \}^{-1} du \;\; \; (\forall x \in (0,+\infty)) .
\end{equation}

At equiprobability (i.e., $p_i=1/W$) we have

\begin{equation}
S_{\{f\}} =   \int_1^W \left \{ \int_{-\infty}^{+\infty} f(\tau) \,u^\tau d\tau\right \}^{-1} du
\end{equation}

\section{Properties of the generalized logarithm and exponential functions and the corresponding entropy}

We list now some useful properties that can be easily proven for any {\it normalized}  distribution $f(q)$, given the fact that $f(q)$ is nonnegative.
\begin{equation}
\ln_{\{f\}} 1=0 \;,
\end{equation}
hence
\begin{equation}
\exp_{\{f\}} 0=1\;.
\end{equation}
Also
\begin{equation}
\left . \frac{d}{dx} \ln_{\{f\}} x \right |_{x=1} =    
\left . \frac{d}{dx} \exp_{\{f\}} x \right |_{x=0} =  1 \;,
\end{equation}
as well as {\it monotonicity}, more precisely
\begin{equation}
\frac{d}{dx} \ln_{\{f\}} x > 0, \; \forall x \in (0,+\infty) \;,
\end{equation}
and
\begin{equation}
\frac{d}{dx} \exp_{\{f\}} x > 0, \; \forall x \in \mathcal{A}_{exp_{\{f\}}} \;,
\end{equation}
where $\in \mathcal{A}_{exp_{\{f\}}}$ is the set of admissible values of $x$ for the nonnegative $\exp_{\{f\}} x $ function. When no negative $q$ contributes (i.e., if $f(q)=0,\; \forall q<0$), then the following properties hold also:

\begin{equation}
\frac{d^2}{dx^2} \ln_{\{f\}} x < 0
\,\,\,\, \text{({\it concavity})} \, ,
\end{equation}
and
\begin{equation}
\frac{d^2}{dx^2} \exp_{\{f\}} x < 0
\,\,\,\, \text{({\it convexity})} \, .
\end{equation}

Analogously, when no positive $q$ contributes (i.e., if $f(q)=0,\; \forall q>0$), then
\begin{equation}
\frac{d^2}{dx^2} \ln_{\{f\}} x > 0
\,\,\,\, \text{({\it convexity})} \, ,
\end{equation}
and
\begin{equation}
\frac{d^2}{dx^2} \exp_{\{f\}} x > 0
\,\,\,\, \text{({\it concavity})} \, .
\end{equation}

Also, if no $q$ above unity contributes (i.e., if $f(q)=0,\; \forall q>1$), then

\begin{equation}
\lim\limits_{x\rightarrow +\infty} \ln_{\{f\}} x=+\infty \;,
\end{equation}

and, if no $q$ below unity contributes (i.e., if $f(q)=0,\; \forall q<1$), then

\begin{equation}
\lim\limits_{x\rightarrow 0^+} \ln_{\{f\}} x=-\infty \;.
\end{equation}

\section{Examples of the generalized entropy}

\subsection{Gaussian distribution}

As a first example of QSF, let us consider a Gaussian distribution of entropic indices around
a central index $q$ with a variance $\sigma>0$:

\begin{equation}
f(\tau)=\frac{1}{\sigma \sqrt{2\pi}} \, e^{-\frac{(\tau-q)^2}{2\sigma^2}}
\end{equation}

Using equation \eqref{eqln} we obtain, for the corresponding generalized logarithm:

\begin{equation}
\ln_{\{q,\sigma\}}x=\frac{\sqrt{\pi}}{\sqrt{2}\sigma} \,
e^{\frac{(1-q)^2}{2\sigma^2}} 
\left [ erf \left ( \frac{1-q}{\sqrt{2}\sigma}
\right ) - erf \left (\frac{1-q-\sigma^2 \ln x}{\sqrt{2}\sigma} \right ) \right ]
\end{equation}


Where $erf$ is the error function. As special cases we obtain:
\begin{equation}
\ln_{\{q,0\}}x=\frac{x^{1-q}-1}{1-q}\;,
\end{equation}
and
\begin{equation}
\ln_{\{1,\sigma\}}x=\frac{\sqrt{\pi}}{\sqrt{2}\sigma} \,  erf \left
(\frac{\sigma \ln x}{\sqrt{2}} \right )
\end{equation}

The entropy functional is given by

\begin{equation}\label{entropyG}
S_{\{q,\sigma\}}=\frac{\sqrt{\pi}}{\sqrt{2}\sigma} \, e^{\frac{(1-q)^2}{2\sigma^2}} 
\sum_{i=1}^W p_i \left [ erf \left ( \frac{1-q}{\sqrt{2}\sigma} \right ) - erf
\left (\frac{1-q+\sigma^2 \ln (1/p_i)}{\sqrt{2}\sigma} \right ) \right ]
\end{equation}

In Fig. 1 we present, for typical values of $(q,\sigma)$, the entropic function $s_{\{q,\sigma\}}(p)$, the entropy $S_{\{q,\sigma\}}(p)$ for a two-state system, and the microcanonical entropy $S_{\{q,\sigma\}}(W)$ as a function of the number of states $W$.

\subsection{Two entropic indices}

The general problem of two indices in a QSF corresponds to the form:

\begin{equation}\label{gentwo}
f(\tau)=a_{1} \delta(\tau-q_1) + a_{2} \delta(\tau-q_2) 
\end{equation}

(If we wish to have a normalized QSF we need to require $a_1+a_2=1$.) Since the general case of Eq. (30) is quite untractable, we will focus our attention onto two special cases, referred to as the {\it symmetric} and the {\it asymmetric} ones. Let us first consider the case where the two indices are symmetric with regard to
$q=1$, i.e., $q_1=1-L/2$ and $q_2=1+L/2$ and $a_1=a_2=1/2$. We then have:

\begin{equation}
f(\tau)=\frac{1}{2} \left [ \delta (\tau-1+L/2) + \delta (\tau-1-L/2) \right ]
\end{equation}
hence, using \eqref{eqln}, 

\begin{equation}
\ln_L x=\int_{1}^x \frac{2}{u^{1-L/2} + u^{1+L/2}} du 
=\frac{4 \, arctan(x^{L/2})-\pi}{L}
\end{equation}

In the limit $L \rightarrow 0^+$ we recover $\ln_L x=\ln x$ and we will thus be
driven back to the BG case. The entropy functional is given by:

\begin{equation}
S_L=\sum_{i=1}^W p_i \, \frac{4 \, arctan(p_i^{-L/2})-\pi}{L}
\end{equation}

It can be readily seen that, for equiprobability, the entropic functional is given by
\begin{equation}\label{microSE}
S_L=\frac{4\, arctan(W^{L/2})-\pi}{L}
\end{equation}
Notice that, when $W \rightarrow \infty$,
the entropy $S_L$ remains finite, in contrast with the BG case.

In Fig. 2 we present, for typical values of $L$, the entropic function $s_L(p)$, the entropy
$S_L(p)$ for a two-state system, and the microcanonical entropy $S_L(W)$ as a function of the number of states $W$.

We consider now the asymmetric case, defined as the special case of Eq. \eqref{gentwo} where one of the indices equals unity.
Then we have:

\begin{equation}
f(\tau)=a_1 \delta(\tau -1) + a_q \delta(\tau-q) \;\;\;\;(a_1+a_q=1)
\end{equation}

This form might be of physical relevance. As already mentioned, unless $a_1+a_q=1$, $f(\tau)$ is {\it not} normalized (see \cite{remark}). The differential equation corresponding to the $(a_1,a_q)$ generic case has already been successfully used (in order to describe, for instance, the re-association of folded heme proteins \cite{TsallisBemskiMendes99} and the flux of cosmic rays \cite{TsallisAnjosBorges03}). 

It can be proven that the corresponding generalized logarithm is:

\begin{equation}
ln_{\{a_1,a_q,q\}}x=\frac{1}{a_1(1-q)} \ln \left [ 1-\frac{a_1}{a_1+a_q}\left ( 1-x^{1-q} \right ) \right ]
\end{equation}
We straightforwardly verify that $\ln_{\{a_1,0,q\}}x = \ln x$ and that $\ln_{\{0,1,q\}}x =\ln_q x$. 
The corresponding entropic functional can be writen as:

\begin{equation}
S_{\{a_1,a_q,q\}}=\frac{1}{a_1(1-q)} \sum\limits_{i=1}^W p_i \ln \left [ 1-\frac{a_1}{a_1+a_q}\left ( 1-p_i^{q-1} \right ) \right ]
\end{equation}

In the microcanonical case this can be writen as:

\begin{equation}
S_{\{a_1,a_q,q\}}=\frac{1}{a_1(1-q)} \ln \left [ 1-\frac{a_1}{a_1+a_q}\left ( 1-W^{1-q} \right ) \right ]
\end{equation}

Like in the previous case, it can be shown that, for $q>1$, the entropy remains finite when $W\rightarrow +\infty$. In Fig. 3 we present, for typical values of $(q,a_q)$ and $a_1=1-a_q$, the entropic function $s_{\{a_1,a_q,q\}}(p)$, the entropy $S_{\{a_1,a_q,q\}}(p)$ for two states, and  the entropy $S_{\{a_1,a_q,q\}}(W)$ in the microcanonical ensemble as a function of the number of states $W$.

\section{Calculation of the QSF for a given entropic functional}

We have seen so far how from a given QSF we can produce the corresponding
entropic functional. We will now work in the reverse way: given a specific entropic
functional, we will find (if possible) the QSF that produces it. Consider
a general entropic functional of the form:

\begin{subequations}
\begin{equation}
S=\sum_{i=1}^W s(p_i)
\end{equation}
\begin{equation}
s(x)=x\,\ln_{\{f\}}\frac{1}{x}
\end{equation}
\end{subequations}

We have:

\begin{equation*}
\ln_{\{f\}}x=\int_1^x \frac{du}{\int_{-\infty}^{+\infty}f(\tau)u^\tau d \tau}
\Leftrightarrow
\end{equation*}
\begin{equation*}
\frac{d}{dx}\left ( \ln_{\{f\}}x
\right ) =\frac{1}{\int_{-\infty}^{+\infty}f(\tau)x^\tau d \tau}\Leftrightarrow
\end{equation*}
\begin{equation*}
\int_{-\infty}^{+\infty}f(\tau)x^\tau d
\tau=\frac{1}{\frac{d}{dx}\left ( \ln_{\{f\}}x \right )}\Leftrightarrow
\end{equation*}
\begin{equation}\label{eq1}
\int_{-\infty}^{+\infty}f(\tau)e^{\tau \, \ln x} d
\tau=\frac{1}{\frac{d}{dx}\left ( \ln_{\{f\}}x \right )}
\end{equation}

We set:

\begin{equation}
\omega=-i\, \ln x
\end{equation}

Then, from equation \eqref{eq1}, we obtain:

\begin{equation}\label{eq2}
\frac{1}{\sqrt{2\pi}}\int_{-\infty}^{+\infty}f(\tau)e^{i\omega \tau} d
\tau=\frac{1}{\sqrt{2\pi}}\cdot \frac{e^{i\omega}}{\frac{d}{d\omega}\left (
\ln_{\{f\}}e^{i\omega} \right )}
\end{equation}

The LHS of equation \eqref{eq2} is however nothing but the Fourier transform of
$f$. Thus, inverting the transform we have:

\begin{equation}\label{eq3}
f(q)=\frac{i}{2\pi}\int_{-\infty}^{+\infty} \frac{e^{i\omega
(1-q)}}{\frac{d}{d\omega}\left ( \ln_{\{f\}}e^{i\omega} \right )}\, d\omega
\end{equation}

Inserting the entropy functional of Eq. (7) into equation \eqref{eq3} we finally get:

\begin{equation}\label{eq4}
f(q)=\frac{1}{2\pi}\int_{-\infty}^{+\infty} \frac{e^{-i\omega q}}
{s\left( e^{-i\omega} \right ) -i\, \frac{d}{d\omega} \left [
s\left( e^{-i\omega} \right ) \right ]}\, d\omega
\end{equation}

Equation \eqref{eq4} is quite important. Indeed, it shows that the QSF corresponding to a
large class of entropy functionals can be explicitly calculated. It is straightforward to check that for $s$ given by BG or by nonextensive statistics we
get $f(q)=\delta(q-q_0)$ as anticipated.

In order to be able to derive the normalized QSF associated with a given entropy,
the entropy functional must fulfill the following requirements.

\begin{enumerate}
\item
It must be possible to write the total entropy $S$ as a sum of the entropic function $s$ for each
state (Eq. (39a)).
\item
The function $s$ must satisfy $s(1)=0$, which is in fact a quite reasonable requirement for an entropy.
\item
Furthermore, we must have:

\begin{equation}
\left . \frac{ds(x)}{dx} \right |_{x=1} = -1
\end{equation}

This condition is equivalent to having the QSF normalized to unity. If we
abandon the normalization of the QSF then we can consistently drop this last
requirement.
\item
The function $s(x)$ must be defined (or analytically continued) on the unitary circle
and it must also be differentiable in the same domain.
\item
The integral of Eq. \eqref{eq4} must converge.
\end{enumerate}

As a nontrivial illustration, we will now use the present method to find the QSF associated with an exponential entropic form. Let us assume 

\begin{equation}
S=\sum_{i=1}^W p_i\left(1-e^{\frac{p_i-1}{p_i}} \right) \,,
\end{equation}
hence

\begin{equation}\label{entexp}
s(x)=x\left(1-e^{\frac{x-1}{x}} \right ) \,.
\end{equation}

It is trivial to see that Eq.\eqref{entexp} fulfills all the criteria set above, and we can
thus find a normalized QSF for it. Using Eq.\eqref{eq4} we get:

\begin{equation}
f(q)=\frac{1}{e}\, \sum_{n=0}^{\infty} \frac{\delta(q-n)}{n!}
\end{equation}
Although different, entropy (46) has some resemblance with that introduced by Curado \cite{Curado}. We claim no particular physical justification for the form (46). In the present context, it has been chosen uniquely with the purpose of illustrating the mathematical procedure involved in the inverse QSF problem.

\section{The canonical ensemble}

To obtain, for the canonical ensemble within the standard nonextensive thermostatistics, the distribution corresponding to the physically relevant stationary state, we must optimize the entropic functional

\begin{equation}
S_q=\sum_{i=1}^W p_i \ln_q \frac{1}{p_i}  \,.
\end{equation}
Besides normalization of the probabilities, we use the following energy constraint (see \cite{TsallisMendesPlastino98,AbeShoreJohnson} and references therein):
\begin{equation}
\frac{\sum_{i=1}^W p_i^q E_i} {\sum_{i=1}^W p_i^q}=U \,,
\end{equation}
where $\{E_i\}$ are the eigenvalues of the Hamiltonian.
The optimizing distribution can be written as a generalized exponential form as follows:
\begin{equation}
p_i=\frac{\exp_q^{-\beta E_i}}{\sum_{i=1}^W \exp_q^{-\beta E_i}}\,,
\end{equation}
where $\beta$ is an inverse-temperature like parameter. As we see, the stationary distribution involves precisely the {\it inverse} function of the generalized logarithm that entered into the definition of the entropy. The following question arises: {\it In the general case considered in  the present paper, is it possible to write the energy constraint in such a way that the stationary state distribution is described by the inverse function of the generalized logarithm used in the definition of the entropy?} Let us right away say that, if this is possible, we have not succeded in doing it. It seems that this nice property is not very general, and it appears to be lost when we extend  further (or differently) than nonextensive thermostatistics.

Let us present an attempt along that line. We assume that the entropy is given by Eq. (7).
The corresponding energy constraint is assumed to be given as follows:
\begin{equation}
\frac{\sum_{i=1}^W \int_{-\infty}^{+\infty} f(q) p_i^q dq E_i} {\sum_{i=1}^W \int_{-\infty}^{+\infty} f(q) p_i^q dq}=U \,.
\end{equation}
In the case where $f(q)=\delta(q-q_0)$ we get back to the usual nonextensive constraint. 

The stationary state canonical probability distribution is the one that optimizes the entropy functional
under this energy constraint. In all the examples that we have given in Section IV, this calculation is analytically untractable. Nevertheless, it can be trivially shown that the solution \emph{is not} the inverse function of the generalized logarithm that entered into the definition of the entropy.
It seems thus, that the exponential form
is not a generic feature but rather a peculiar ``coincidence'', only valid for the nonextensive family of entropies $S_q$. It remains, however, as an open and rather challenging problem to come up with an energy constraint that would produce the corresponding generalized exponential in the generic case. 

\section{Connection with the Beck-Cohen superstatistics}

Recently, a generalization of the BG statistics has been proposed by Beck and Cohen. This
generalized superstatistics assumes that there may be strong deviations from the classical BG
case when there are temperature fluctuations within the system. Along that line, the Boltzmann factor
$\exp(-\beta E)$ is generalized as follows:

\begin{equation}
B(E)=\int_0^{+\infty} g(\beta) e^{-\beta E} d\beta
\end{equation}

Where $g(\beta)$ is the probability distribution of the inverse tempature in the system.
Quite recently, an entropic functional has been derived that corresponds to superstatistics.
This functional is of the form:

\begin{subequations}\label{bc}
\begin{equation}
S=\sum_i s(p_i)
\end{equation}
\begin{equation}\label{bc2}
s(y)=\int_0^x \frac{a+K^{-1}(y)}{1-\frac{K^{-1}(y)}{E^*}}
\end{equation}
\begin{equation}
\text{where } K(y)=\frac{B(y)}{\int_0^{+\infty} B(u)du}
\end{equation}
\end{subequations}

In Eqs. \eqref{bc}, $E^*$ stands for the lowest admissible energy for the system and $a$ is a Lagrange multiplier.
Using Eq. \eqref{bc2} and Eq. \eqref{eq4} we can in principle get the QSF $f(q)$ for a given temperature
distribution function $g(\beta)$. Thus, in principle at least, the various superstatistics can be accomodated into
the presently proposed formalism.

There are certain cases however, where the present formalism can go further. For example, it appears that through
superstatistics we can only produce the so-called nonextensive entropies $S_q$ for $q \ge1$, while in the present formalism we can have them for
arbitrary values of $q$.

\section{Conclusions}

In the present work, we study the properties of a generalized entropic function based on a
distribution of $q$ entropic indices. We introduce the idea of $q$-spectral function(QSF) and
exhibit three nontrivial examples: the case of a Gaussian, as well as both symmetric and nonsymmetric choices within a sum of two delta
functions. We also introduce the notion of an inverse QSF problem, in the sense that, given some entropic functional,
we present a procedure for explicitly calculating the QSF that produces it.

Also, we have briefly discussed the canonical ensemble and its corresponding energy constraint. We have come to the rather unexpected conclusion that, 
in the general case, the stationary state probability distribution seems to be \emph{not} the generalized exponential which is the {\it inverse} function of the generalized logarithm used to define the entropy. Altering the form
of the constraint in order to produce such exponential remains therefore as an open problem.  

We have also pointed out the connection of the present work to the Beck-Cohen superstatistics. Incidentally, we have shown that there are examples of entropies that can be produced within the current theory which appear to be not admissible within the frame of Beck-Cohen superstatistics.

It would certainly be very interesting to apply the present formalism to concrete physical systems (e.g., high precision experiments in turbulent systems) that would defy more conventional approaches.  On a different vein, it would be theoretically valuable to find nontrivial special cases for which the canonical ensemble 
distribution could be explicitly calculated, since this would allow a deeper analysis of the properties of the optimizing distribution.

\begin{acknowledgments}
The authors would like to gratefully acknowledge fruitful and enlightening discussions with Dr. A. Provata.
\end{acknowledgments}

\clearpage


\begin{figure}[htbp]
\begin{center}
\hspace{3mm}
\includegraphics[scale=1.3]{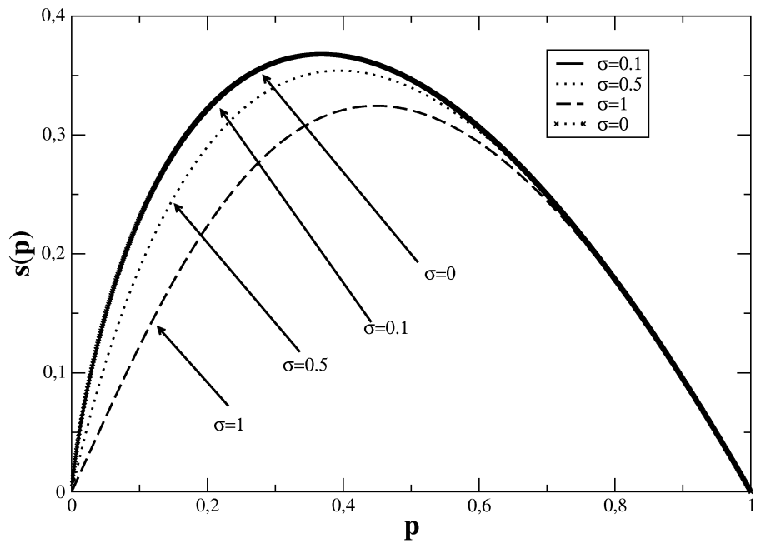}
\end{center}

\begin{center}
\includegraphics[scale=1.3]{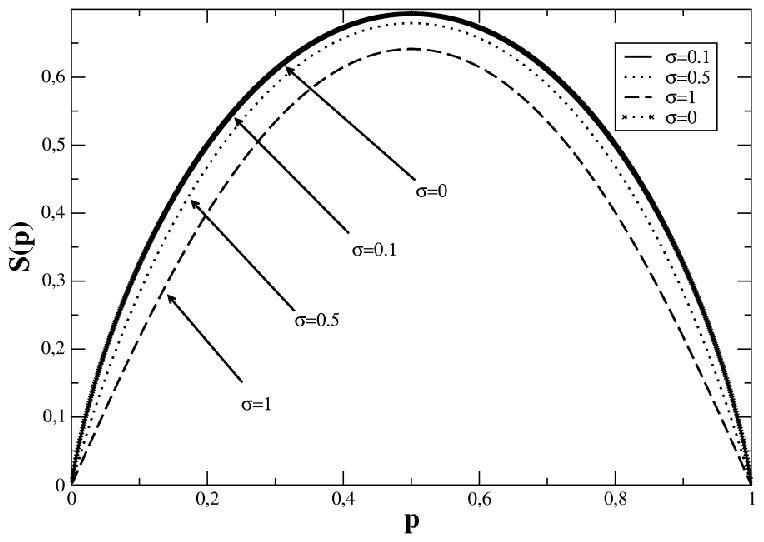}
\end{center}

\begin{center}
\includegraphics[scale=1.3]{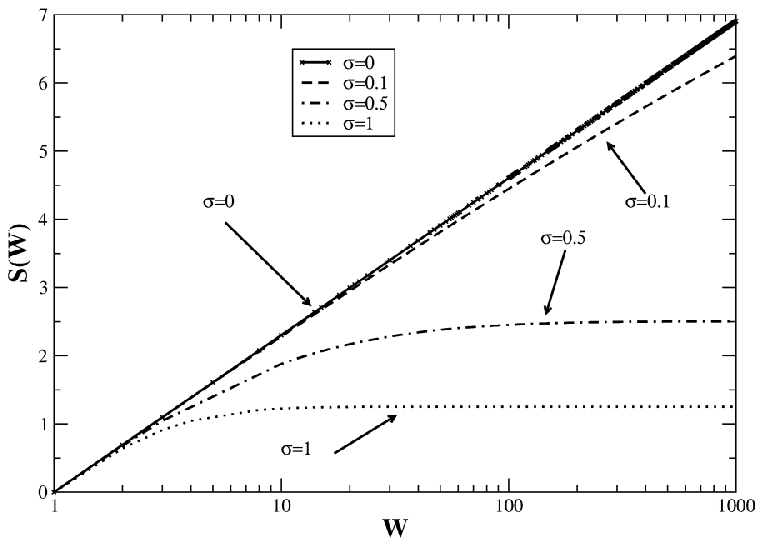}
\end{center}
\caption{The entropic function $s_{\{q,\sigma\}}(p)$ ({\it top}),   the entropy $S_{\{q,\sigma\}}(p)$ for a two-state system   ({\it middle}), and  the $W$-state microcanonical-ensemble entropy $S_{\{q,\sigma\}}(W)$  ({\it bottom}). The four lines correspond to $q=1$ and $\sigma=0.1$, $\sigma=0.5$, $\sigma=1$ and $\sigma=0$ (BG case).
}
\end{figure} 

\begin{figure}[htbp]
\begin{center}
\hspace{3mm}
\includegraphics[scale=1.3]{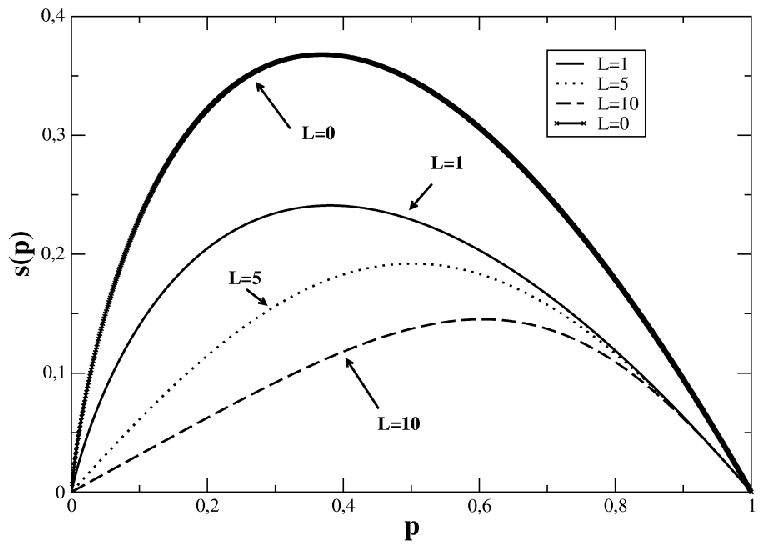}
\end{center}

\begin{center}
\includegraphics[scale=1.3]{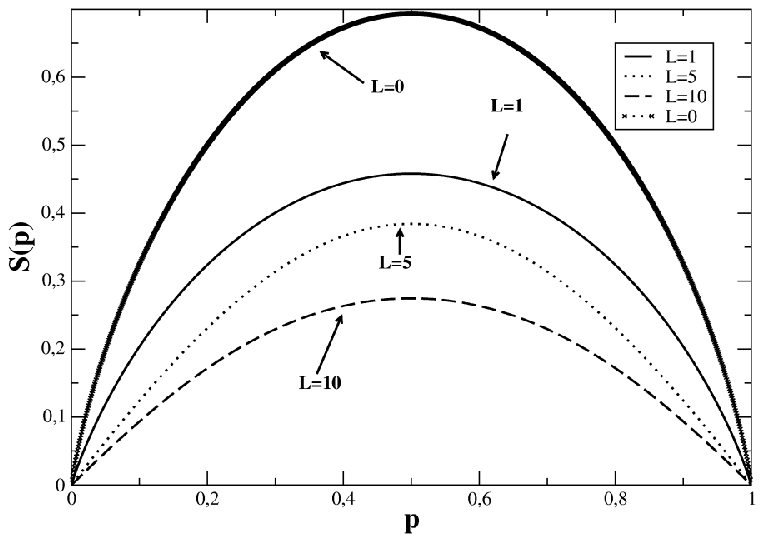}
\end{center}

\begin{center}
\includegraphics[scale=1.3]{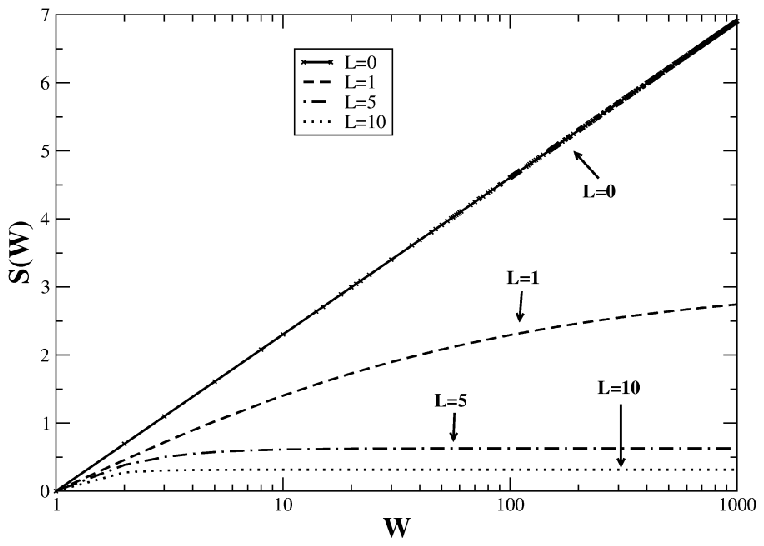}
\end{center}
\caption{The entropic function  $s_L(p)$          ({\it top}),    the entropy $S_L(p)$ for a two-state system   ({\it middle}), and  the $W$-state microcanonical-ensemble entropy $S_L(W)$  ({\it bottom}). The four lines correspond to $L=1$ , $L=5$, $L=10$ and for $L=0$ (BG case).
}
\end{figure} 

\begin{figure}[htbp]
\begin{center}
\hspace{3mm}
\includegraphics[scale=1.3]{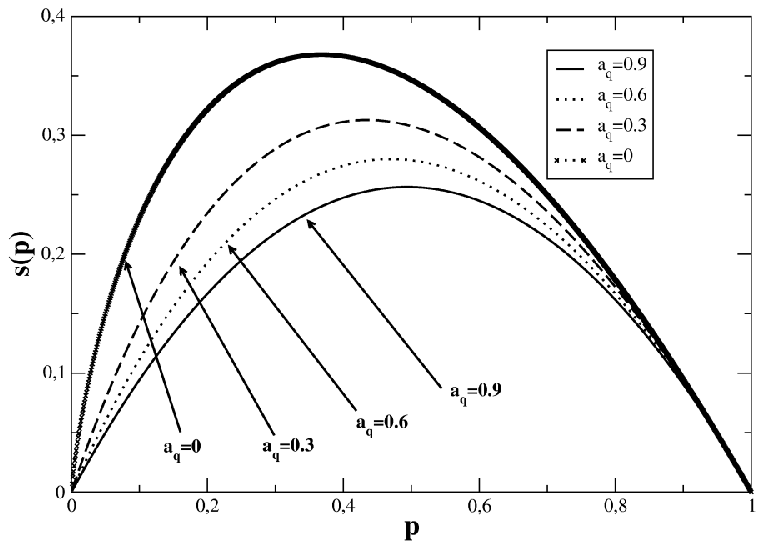}
\end{center}

\begin{center}
\includegraphics[scale=1.3]{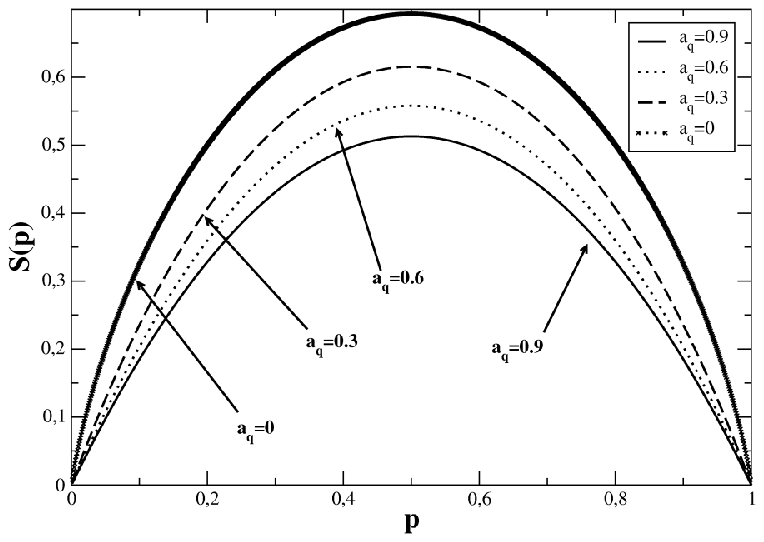}
\end{center}

\begin{center}
\includegraphics[scale=1.3]{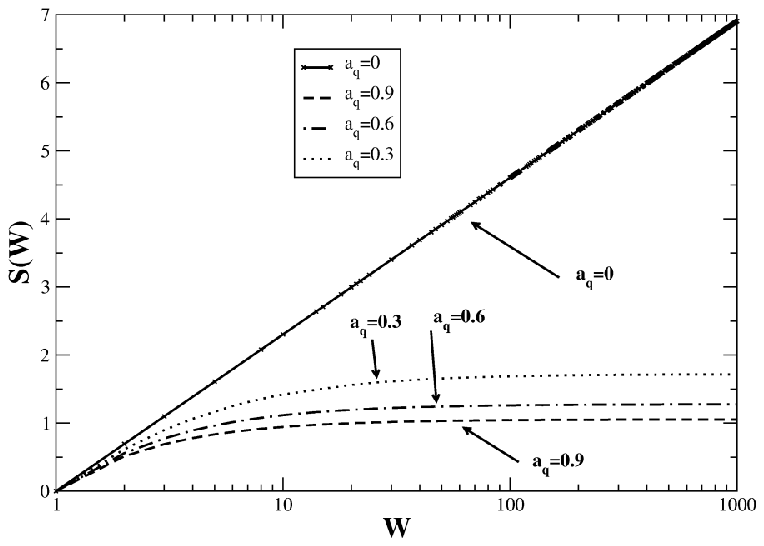}
\end{center}
\caption{The entropic function $s_{\{a_1,a_q,q\}}(p)$ ({\it top}),       the entropy $S_{\{a_1,a_q,q\}}(p)$ for a two-state system   ({\it middle}), and  the $W$-state microcanonical-ensemble entropy $S_{\{a_1,a_q,q\}}(W)$  ({\it bottom}). The four lines correspond to $q=2$, $a_1=1-a_q$, and $a_q=0.9$, $a_q=0.6$, $a_q=0.3$ and $a_q=0$ (BG case). 
}
\end{figure} 


\begin{thebibliography}{10}
\bibitem{remark}Let us assume that $f(\tau)$ is such that $\int_{-\infty}^{+\infty} f(\tau) \,d\tau=A$, with $|A|< \infty$,  $A$  not necessarily being unity. We define $F(\tau) \equiv f(\tau)/A$, hence $\int_{-\infty}^{+\infty} F(\tau) \,d\tau=1$. We then have that $\ln_F(x)=\int_1^x \left \{ \int_{-\infty}^{+\infty} F(\tau) \,u^\tau d\tau
\right \}^{-1} du =A \ln_f(x) $. For example, in the case of Eq. (36), we can verify that 
$\ln_{\{\frac{a_1}{a_1+a_q},\frac{a_q}{a_1+a_q}\}} x =(a_1+a_q) \ln_{\{a_1,a_q\}} x$.
\end{thebibliography}
\end{document}